\documentclass[final, times, twocolumn]{elsarticle}

\usepackage{graphicx}
\graphicspath{ {./figures/} }
\usepackage{multirow}
\usepackage{amsmath}
\usepackage{xcolor}
\usepackage{soul} 
\usepackage[para]{threeparttable}
\usepackage{comment}
\usepackage{siunitx}
\usepackage{mathtools}

\usepackage{hyperref}

\journal{International Journal of Heat and Mass Transfer}

\begin{document}

\begin{frontmatter}

\title{Optical Absorption Effects in Thermal Radiation Barrier Coating Materials}

\author{Georgios Koutsakis\corref{mycorrespondingauthor}}

\cortext[mymainaddress]{Corresponding author}
\ead{koutsakis@seas.harvard.edu}

\author{David R. Clarke}

\address{School of Engineering and Applied Sciences, Harvard University, Cambridge MA, 02138, USA}

\begin{abstract}

    Future gas turbine engines will operate at higher gas temperatures and consequentially hot-section components such as blades, vanes and combustors, will be subject to higher thermal radiation fluxes than today. Current thermal barrier coating materials are translucent over the spectral region of the heat flux so future coatings will also have to provide a barrier to thermal radiation. The effects of optical absorption and scattering properties of coating materials on the temperatures and heat fluxes through coatings are explored using a two-flux heat transfer model, and promising combinations are identified that reduce the coating-alloy interface temperatures. Lower interface temperatures occur for thickness normalized absorptions of $\overline{\kappa} L$ $>$1. The effect of both a narrow and a broad band spectrally selective absorbing Gd${_2}$Zr${_2}$O$_{7}$ based coating materials are then studied. These show that large values of the product of the normalized absorption length and the spectral width of the absorption are required to significantly decrease the radiative heat transport through a coating. The results emphasize the importance of enhancing the optical absorption of the next generation barrier materials as a strategy to increase gas turbine engine efficiency by decreasing compressor bleed air cooling requirements.
    
\end{abstract}


\end{frontmatter}

\newpage

\section{Introduction}
	
\begin{figure}[!htbp]
		\centering
		\includegraphics[width=1\columnwidth]{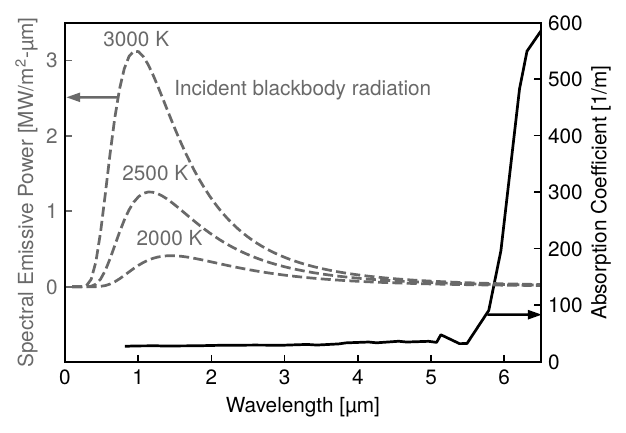}
		\caption{Comparison between the blackbody spectral emissive power (dashed, right axis) for the indicated temperatures, and the absorption in the visible/infrared region of a Gd${_2}$Zr${_2}$O$_{7}$ plasma-sprayed thermal barrier coating~\cite{wang2013thermal}}
		\label{fig:optical-gap-blackbody}
	\end{figure}

	Thermal barrier coatings are used to provide thermal protection to metal alloy components in gas turbines so that, in combination with internal cooling, the alloy surface temperature does not exceed a maximum, materials dependent, value. Since their introduction, in the 1980s, the coating materials have been selected based on having a low thermal conductivity. However, as combustion temperatures in gas turbines increase with the development of more efficient engines, the radiative heat loading on thermal barrier coatings are expected to increase. This raises important questions about the materials selection for coatings, coating thicknesses and possibly, the additional cooling required to keep the temperatures within the capabilities of various coating and alloys.  To address these concerns, it is necessary to determine how much of the thermal radiation propagates through thermal barrier coatings, how the temperature distribution through a coating is modified by the radiative component in the coating and how the optical properties of a coating might be changed in order to alter the temperature distributions through the coating as well as the heat fluxes reaching the interface between the coating and the underlying metal alloy.  These are important questions as current thermal barrier coatings, such as yttria-stabilized zirconia (YSZ) and gadolinium zirconate (GZO), are translucent and exhibit little absorption over the spectral range over which a black body radiates at current and anticipated future gas temperatures. This ``thermal radiation window'' is illustrated in Fig.~\ref{fig:optical-gap-blackbody} which compares the blackbody spectral emissive power (right) for different gas temperatures, and the spectral absorption coefficients for a commercial Gd${_2}$Zr${_2}$O$_{7}$ thermal barrier coating. As indicated, the coating exhibits little or no absorption from about 0.5 \si{\micro\metre} to 6.5 \si{\micro\metre}. Similarly, YSZ also exhibits similar, low absorption up to about 6.5 \si{\micro\metre} ~\cite{wahiduzzaman1989effects, eldridge2009determination, wang2013thermal}.
    
    In modeling heat transfer through coatings, it is usually assumed that heat transfer is based on thermal conduction alone although the original work by Siegel \cite{siegel1996internal,siegel1997green,siegel1998analysis}, as well as some recent studies have discussed the possible contributions of radiative transport~\cite{flamant2019opportunities, johnson2021analysis}. In this work we extend these studies, based on continuum, two-flux heat transfer models, to parameterize the temperature and flux distributions through thermal barrier coatings, and discuss the role of optical absorption and scattering within the coating. We also consider how the heat transfer would be affected by rare earth element doping of a Gd${_2}$Zr${_2}$O$_{7}$ coating, representative electronic dopants that absorbs in the near-infra-red.  The results presented are then used as the basis for discussing the effects of different ``grey'' absorption bands. As heat transfer through coatings is dependent on many different parameters, the calculations presented here are specifically for Gd${_2}$Zr${_2}$O$_{7}$ coatings, deposited on a nickel-based superalloy with internal cooling, subject to the same boundary conditions as have previously been used in similar computations \cite{siegel1998analysis,flamant2019opportunities}. These will be described in the following section in which the spectral two-flux model is summarized and how it was implemented. 
	
\section{The Spectral Two-Flux Model}
	
\begin{figure}[!htbp]
		\centering
		\includegraphics[width=1\columnwidth]{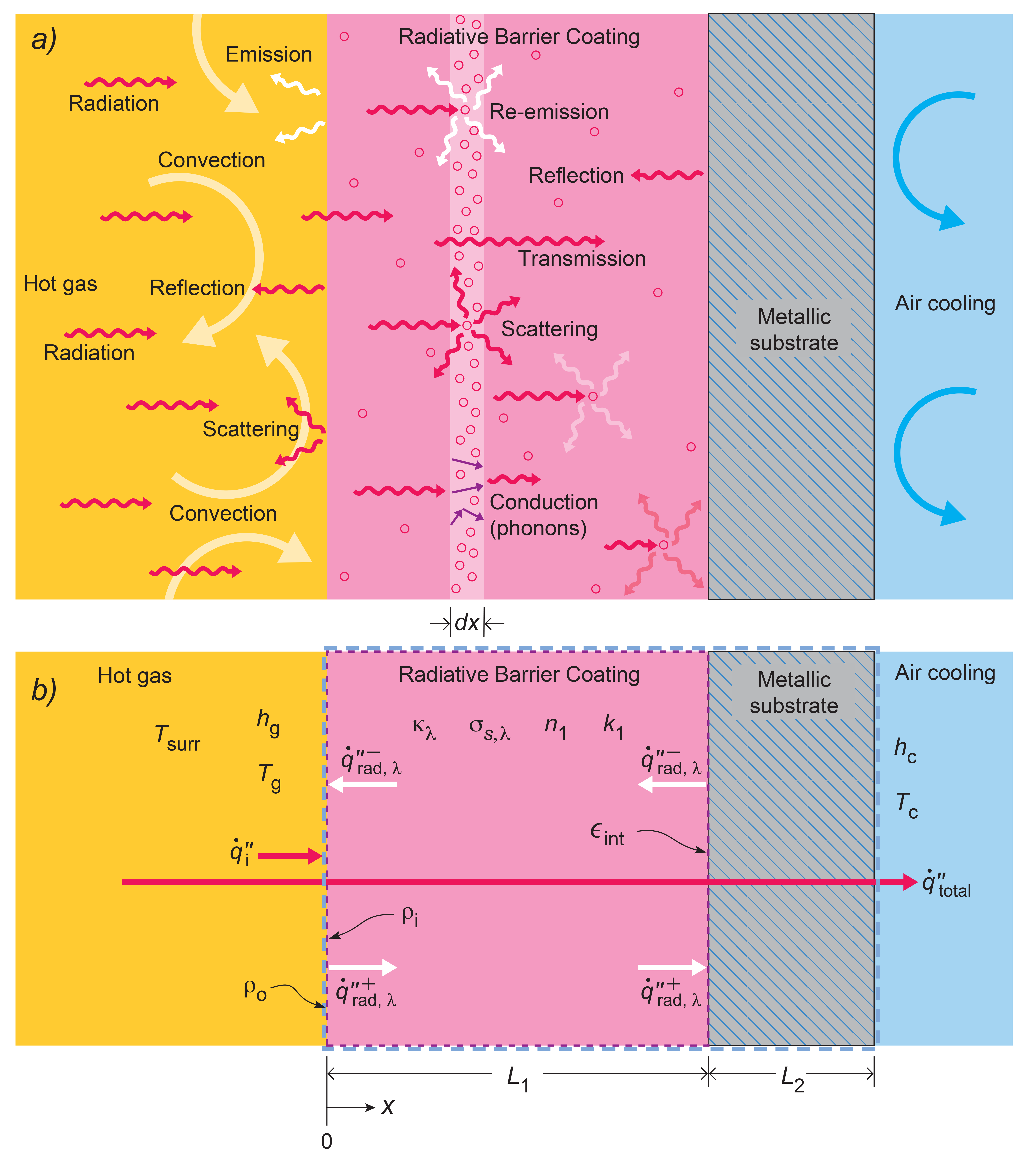}
		\caption{(a) Schematic illustration of the heat transfer mechanisms when a translucent thermal barrier coating on an internally cooled alloy gas turbine engine blade is heated by a hot gas and heat extracted by the internal cooling. In propagating through the coating, the thermal radiation can be absorbed, scattered, re-emitted and or converted to phonons. (b) The geometry and nomenclature used in the two-flux model to describe the processes shown in (a). The heat fluxes through the coating are the sum of the conductive and radiative heat fluxes, evaluated self-consistently.}
		\label{fig:sketch}
	\end{figure}

	In this section, we summarize the essential features of the spectral two-flux model introduced by \cite{siegel1996internal, siegel1997green, siegel1998analysis} to analyze the combination of radiative and conductive heat transport through a coating on an internally cooled alloy component. The individual physical processes are shown schematically in Fig.~\ref{fig:sketch} (a) which is also used to introduce the terms used in the two-flux model. The outer surface of the coating is exposed to a hot gas thermally radiating into the coating material. In the one-dimensional representation, it is useful to consider that the coating consists of layers of incremental thickness which absorb, scatter, transmit and re-emit radiation as well as resisting thermal conductance. It is assumed that the absorption coefficient of the incremental layer is numerically equal to its emissivity. Also, the absorption and scattering coefficients are assumed, in our implementation of the two-flux model, to be wavelength dependent.
    Figure~\ref{fig:sketch} (b) illustrates the energy and radiative transfer balances within the dashed and dotted control volumes, respectively. A steady state condition is assumed so that the total heat flux is independent through the thickness of the coating and the underlying alloy.
	In the hot gas the heat flux equals to sum of convection and blackbody radiation, in the coating equals to the sum of conduction and radiation, in the metal alloy equals only to conduction and in the cooling surface it equals only to convection.
	Within the translucent coating layer the energy balance requires that the total heat flux, $\dot{q}^{\prime \prime}_{\text{total}}$, equals the sum of the conductive, $\dot{q}^{\prime \prime}_{\text{cond}}$, and radiative, $\dot{q}^{\prime \prime}_{\text{rad}}$, fluxes, and at steady state the total heat flux is independent of depth into the coating, $x$.
	The conduction and radiation fluxes depend on depth as:
	\begin{equation}\label{energybalancefunctionofx}
		\begin{split}
			\dot{q}^{\prime \prime}_{\text{total}} & = \dot{q}^{\prime \prime}_{\text{cond}} + \dot{q}^{\prime \prime}_{\text{rad}} \\
			& = - k_1 \frac{\,\mathrm{d}T}{\,\mathrm{d}x}(x) + \int_{\lambda = 0}^{\infty} \dot{q}^{\prime \prime}_{\text{rad, }\lambda} (x) \,\mathrm{d} \lambda
		\end{split}
	\end{equation}
	where $k_1$ is the coating thermal conductivity, $T(x)$ is the temperature distribution as a function of depth $x$, $\lambda$ is the wavelength and $\dot{q}^{\prime \prime}_{\text{rad, }\lambda} = \dot{q}^{\prime \prime +}_{\text{rad, }\lambda} - \dot{q}^{\prime \prime -}_{\text{rad, }\lambda}$ is the spectral net radiative heat flux distribution in the coating. The terms $\dot{q}^{\prime \prime +}_{\text{rad, }\lambda}$ and $\dot{q}^{\prime \prime -}_{\text{rad, }\lambda}$ are the diffuse radiosities or radiative heat fluxes in the positive (inward) and negative (outward) direction defined in Fig.~\ref{fig:sketch} (b), respectively.

	Equation \ref{energybalancefunctionofx} is integrated over $x$ to give the temperature distribution in the coating:
	\begin{equation}
		T(x) = T_{x=0} - \frac{x}{k_1} \dot{q}_{\text{total}}^{\prime \prime} - \frac{1}{3 k_1} \int_{\lambda=0}^{\infty} \frac{G_{\lambda, x=0} -G_{\lambda, x}}{\beta_{\lambda}} d\lambda
	\end{equation}
	where $\beta_{\lambda} = \kappa_\lambda + \sigma_{s,\lambda}$ is the wavelength-dependent extinction coefficient, which is the sum of absorption and scattering coefficients, respectively.
	The net spectral radiative heat flux distribution, $\dot{q}^{\prime \prime}_{\text{rad, }\lambda} (x)$, and the spectral intensity of incident radiation distribution, $G_{\lambda}(x)=2\left(\dot{q}^{\prime \prime +}_{\text{rad, }\lambda} + \dot{q}^{\prime \prime -}_{\text{rad, }\lambda}\right)$, are related via the Milne-Eddington approximation~\cite{siddall1974flux}.
	\begin{equation}\label{qlambdaAtinterface}
		\dot{q}^{\prime \prime}_{\text{rad, }\lambda} (x) = -\frac{1}{3 \beta_\lambda} \frac{\,\mathrm{d} G_\lambda}{\,\mathrm{d} x}
	\end{equation}
	The term $G_{\lambda}(x)$ is obtained by solving a second-order differential equation~\cite{siegel1996internal, siddall1974flux} that involves the solid blackbody thermal emission, $n_1^2 E_{\lambda b, T}(x)$.
	\begin{equation}\label{differentialEqforG}
		\frac{\,\mathrm{d}^2 G_{\lambda}(x)}{\,\mathrm{d}x^2}
		- 3 \beta_\lambda^2 \left(1 - \omega_\lambda \right) G_{\lambda}(x) 
		=	
		- 3 \beta_\lambda^2 \left(1 - \omega_\lambda \right) 4 n_1^2 E_{\lambda b, T}(x)	
	\end{equation}
	where $\omega_\lambda$=$\sigma_{s,\lambda}/\beta_{\lambda}$ is the scattering albedo, $n_1$ is the coating refractive index and $E_{\lambda b, T}(x)$=$2 \pi C_1 / \left[n^{2} \lambda^{5} \left(ev^{C_2 / \left(n \lambda T\right)} - 1\right)\right]$ is the blackbody emissive power in which C$_1$ and C$_2$ are the blackbody radiation constants~\cite{howell2020thermal}.
	
	In previous studies, the radiative transport was approximated using a two band approach, to represent a translucent and an opaque band, representative of the optical properties of many high temperature oxides, such as shown in Fig.~\ref{fig:optical-gap-blackbody} with a defined cut-off wavelength above which the coating is opaque. For this case the integral term in Eq.~\ref{energybalancefunctionofx} is considered only within the translucent spectral region, \textit{i.e.} $0 \leq \lambda \leq \lambda_{\text{cut-off}}$. Above the cut-off wavelength, the coating is opaque and no thermal radiation propagates. In implementation of the model in subsequent sections of this contribution, the radiative transport over multiple bands is considered.
	Two boundary conditions are required to solve Eq.~\ref{differentialEqforG}.
    At the coating surface, $x$=0, the radiative heat flux boundary condition is   $\dot{q}^{\prime \prime +}_{\text{rad, }\lambda} = \rho_i \dot{q}^{\prime \prime -}_{\text{rad, }\lambda} + \left(1 - \rho_o \right) \dot{q}_{\text{i}}^{\prime \prime}$.
    This is a mean field approximation to account for the porosity in the coating, where the terms $\rho_i$ and $\rho_o$ represent the inner and outer Fresnel reflectivities for a dielectric, respectively. Their relationships are found elsewhere \cite{howell2020thermal, richmond1963relation, siegel1994approximate}.
	The last term on the right hand side,  $\dot{q}_{\text{i}}^{\prime \prime} = E_{\lambda b} (T_{\text{surr}})$, represents the emissive power of the hot gas, assumed in this section to be a blackbody. 
 
	The other boundary condition, at the coating/alloy substrate interface, $x$=L$_1$, determined as $\dot{q}^{\prime \prime -}_{\text{rad, }\lambda} = \left(1 - \epsilon_{int}\right) \dot{q}^{\prime \prime +}_{\text{rad, }\lambda} + \epsilon_{int} E_{\lambda b}$, where $\epsilon_{int}$ is its emissivity. 
	
	In implementing the model, the two-flux equations for $G_{\lambda}$ and $\dot{q}^{\prime \prime}_{\text{rad, }\lambda}$ are solved for the spectral fluxes in the positive (inward) and negative (outward) directions $\dot{q}^{\prime \prime +}_{\text{rad, }\lambda} = \frac{1}{2} \left(\frac{G_{\lambda}}{2} + \dot{q}^{\prime \prime}_{\text{rad, }\lambda} \right)$ and $\dot{q}^{\prime \prime -}_{\text{rad, }\lambda} = \frac{1}{2} \left(\frac{G_{\lambda}}{2} - \dot{q}^{\prime \prime}_{\text{rad, }\lambda} \right)$, respectively.
	The boundary condition at the coating surface, $x$=0 gives,
	\begin{equation}
		\frac{\,\mathrm{d} G_{\lambda}}{\,\mathrm{d} x}\Biggr|_{\substack{x=0}} = \frac{3}{2} \beta_{\lambda} \frac{1-\rho_i}{1+\rho_i} \left[G_{\lambda (x=0)} - 4 \frac{1-\rho_o}{1-\rho_i}\dot{q}_{\text{i}}^{\prime \prime}\right]
	\end{equation}
	Similarly, the other boundary condition at $x$=L$_1$ yields,	
	\begin{equation}
		\frac{\,\mathrm{d} G_{\lambda}}{\,\mathrm{d} x}\Biggr|_{\substack{x=L_1}} = -\frac{3}{2} \beta_{\lambda} \frac{\epsilon_{int}}{2 - \epsilon_{int}} \left[G_{\lambda (x=L_1)} - 4 n_{1}^2 E_{\lambda b}\right]
	\end{equation}
	where $n_{1}^2 E_{\lambda b}$ is the local blackbody emission $E_{\lambda b} = \sigma T^{4}_{x=L_1}$ with refractive index $n_1$.
	
	The net heat flux energy balance at the coating surface $x=0$ is then the sum of the heat fluxes in the translucent and opaque bands:
	\begin{equation}
		\begin{aligned}
			\dot{q}^{\prime \prime}_{\text{total, translucent}} = &
			\quad h_{\text{g}} \left(T_{\text{g}}-T_{x=0}\right) \\
			& + (1 - \rho_o) \int_{\lambda_{\text{cut-off}}}^{\infty} \left( \dot{q}_{\text{i}}^{\prime \prime} - E_{\lambda b}  \right) \,\mathrm{d} \lambda \\
			& 	- \frac{1}{2} \frac{(1-\rho_i)}{(1+\rho_i)} \int_{0}^{\lambda_{\text{cut-off}}} G_{\lambda, x=0}  \,\mathrm{d} \lambda \\
			&	+ 2 \frac{\left( 1-\rho_o \right)}{\left(1 + \rho_i\right)}  \int_{0}^{\lambda_{\text{cut-off}}}  \dot{q}_{\text{i}}^{\prime \prime} \,\mathrm{d} \lambda 
		\end{aligned}
	\end{equation}
	while for the opaque band, $\lambda > \lambda_{\text{cut-off}}$, the total heat flux occurs only at the surface.
	\begin{equation}
		\begin{aligned}
			\dot{q}^{\prime \prime}_{\text{total, opaque}} = & 
			\quad 	h_{\text{g}} \left(T_{\text{g}}-T_{\text{x=0}}\right) \\
			& \! + \left(1 - \rho_o \right) \sigma \left(T_{\text{g}}^4-T_{\text{x=0}}^4\right)
		\end{aligned}
	\end{equation}

 The total heat flux at the internally cooled passage is purely convective, $\dot{q}^{\prime \prime}_{\text{total, metal}} = \quad h_{\text{c}} \left(T_{x=L_1+L_2}-T_{\text{c}}\right)$. For the purpose of this work an iterative process was carried out to guarantee the total heat flux was constant through the coating and alloy substrate.

	In the calculations presented in this work, several parameters, representing a GZO coating, were held constant unless otherwise stated. The thermal conductivity for the coating and alloy substrate were k$_1$=1.15 W/m-K and k$_2$=33 W/m-K. The refractive index of air was assumed unity, while the coating's effective refractive index was n$_1$=1.58. The outer and inner diffuse reflectivities for the air-coating and coating-air, respectively, were assumed to be hemispherically isotropic, thus, wavelength integrated Fresnel equations~\cite{richmond1963relation} were utilized similar to other studies~\cite{johnson2021analysis, siegel1998analysis}. 
    The emissivity of the alloy and the coating-alloy interface was assumed to be $\epsilon_{int}$=0.3~\cite{siegel1998analysis}. The coating thickness was assumed to be 500 \si{\micro\metre} and the alloy thickness was 762 \si{\micro\metre}. The absorption coefficient, shown in Fig.~\ref{fig:optical-gap-blackbody}, and scattering coefficients deduced from a plasma-sprayed GZO coating can be found elsewhere~\cite{wang2013thermal}.
	The boundary conditions of the hot side included a gas temperature, which was set equal to the blackbody surroundings, at T$_{\mathrm{g}}$=T$_{\mathrm{surr}}$=2500 K and a convective heat transfer coefficient of h$_{\mathrm{g}}$=3014 W/m$^2$K. The assumed hot gas temperature of 2500 K, although somewhat arbitrary, has been adopted to represent the state-of-the-art of future advanced turbines. Although the hot path temperature of today's highest performance aerospace turbines is propriety, it is likely to be lower than 2500 K. Also the current ARPA-E ULTIMATE program for power turbines has a target hot gas inlet temperature of 2073 K, again lower than our assumed gas temperature. (The turbine inlet temperatures of aerospace turbines are generally higher than those of power turbines). In addition, although a high temperature, 2500 K is lower than the melting temperatures of many refractory oxides, including those currently used as thermal barrier coatings. 
	On the cooling side a temperature of T$_{\mathrm{c}}$=1000 K and convective heat transfer coefficient of h$_{\mathrm{c}}$=3768 W/m$^2$K were adopted following Siegel, while the net radiative exchange in the internal cooling passages was neglected~\cite{siegel1998analysis}.
	With these parameters, the boundary value finite difference problem of Eq.~\ref{differentialEqforG} was solved for the radiative component, comprising 100 nodes for 500 \si{\micro\metre} coating and 15 nodes for the alloy substrate.

    \section{Absorption and Scattering Effects}

\begin{figure}[!htbp]
		\centering
		\includegraphics[width=1\columnwidth]{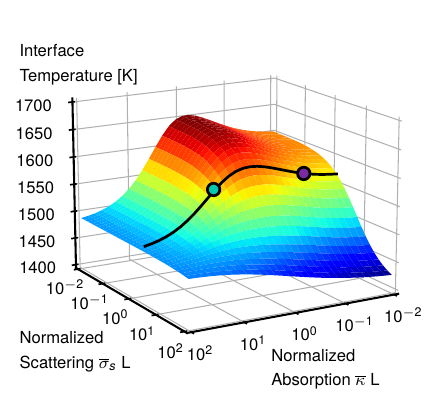}
		\caption{Interface temperature between a coating/substrate system calculated using the two-flux method presented in the text. The terms $\overline{\kappa}L$ and $\overline{\sigma}_s L$ are the normalized absorption and scattering thickness, respectively, and the parameters of the system are as defined in the text. Coating thickness is 500 \si{\micro \meter}. Gas temperature is 2500 K.}
		\label{fig:3d-interface-temperature}
	\end{figure}

	In this section, the two-flux model is used to compute the effects of absorption and scattering on the temperature and heat fluxes in order to identify the optical properties that have the potential to reduce radiative contributions through coatings, such as reducing interface temperatures and heat fluxes. To do this, normalized, wavelength-independent absorption and scattering lengths are introduced as $\overline{\kappa} L$ and $\overline{\sigma}_s L$, respectively, by normalizing by the coating thickness, L. 
    The former is a product of absorption coefficient $\overline{\kappa}$ and coating thickness $L$, and the latter is a product of scattering coefficient $\overline{\sigma}_s$ and coating thickness, L. Large values of $\overline{\kappa} L$ converge to an opaque coating whereas small values tend towards a non-absorbing coating. Unless otherwise stated, the coating thickness used in the computations is 500 \si{\micro\metre}, enabling the computations presented to most clearly show the effects of absorption and scattering of the thermal radiation without simultaneous changes in thermal conduction. 
	Figure~\ref{fig:3d-interface-temperature} shows how the coating-substrate interface temperature depends on the normalized absorption and scattering lengths. 
    For the calculations in this section, the absorption and scattering coefficients were assumed to be spatially uniform throughout the coating. 
	
	Several regimes of behavior are revealed by the results shown in the figure. The effects of absorption are strongest when the absorption length is commensurate with the coating thickness, $\overline{\kappa} L$$\sim$1. At larger values of $\overline{\kappa} L$, the interface temperatures decrease for all values of the scattering coefficient presumably because most of the thermal radiation is absorbed nearer the surface for a coating of fixed thickness. At smaller values of $\overline{\kappa} L$, again for a fixed coating thickness, less thermal radiation is absorbed. A third regime is at high values of the scattering coefficient irrespective of the absorption coefficient.  As will be described later, at these large values of scattering coefficient, the two-flux model breaks down because it does not account in a realistic manner for scattering from microstructural features present in coatings, such as pores and microcracks.  Nevertheless, Fig.~\ref{fig:3d-interface-temperature} emphasizes the importance of optical scattering of thermal radiation.
    In practice the value of the scattering coefficient depends in detail on the process and conditions under which a coating is deposited. Studies on coatings deposited by Plasma Spray and Electro-Beam Physical Vapor Deposition (EB-PVD) reported scattering coefficients of about 10$^3$ m$^{-1}$~\cite{wahiduzzaman1989effects, siegel1998analysis} and roughly about 10$^4$ m$^{-1}$~\cite{limarga2009characterization}, respectively.
	For the purpose of this work, we assume an upper physical limit of normalized scattering length of about $\overline{\sigma}_s$L$\sim$5 in the results presented here.

	Superimposed on Fig.~\ref{fig:3d-interface-temperature} is a contour corresponding to a constant normalized scattering length, $\overline{\sigma}_s$L$\sim$1.6, appropriate to a plasma-sprayed coating, illustrating the variation in interface temperature as the absorption length is systematically altered. 
    There are two points on the contour. The first point (purple) is plasma-sprayed GZO coating~\cite{wang2013thermal}.
    The second point (cyan) corresponds to an equivalent coating with increased normalized absorption coefficient and interface temperature equal to the first.
    Additional discussion for those two cases is provided next.
    
\begin{figure}[!htbp]
		\centering
		\includegraphics[width=1\columnwidth]{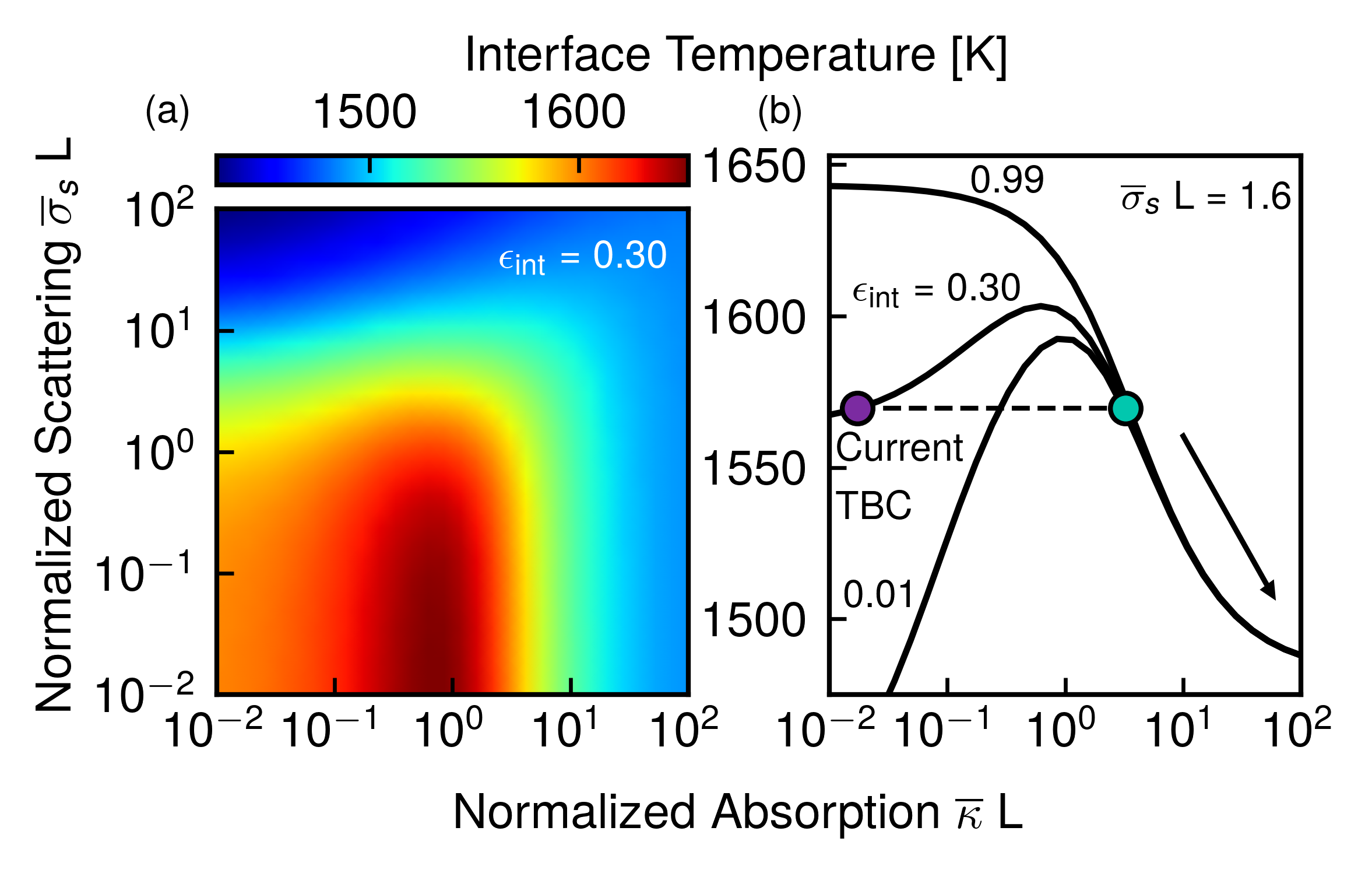}
		\caption{(a) Contours of the coating-substrate interface temperature as a function of the normalized absorption and scattering thickness $\overline{\kappa}$L and $\overline{\sigma}_s$L, respectively. (b) Interface temperature as a function of normalized absorption for a typical normalized scattering is shown for different interfacial emissivities.}
		\label{fig:contour-interface}
	\end{figure}

 
	Figure \ref{fig:contour-interface} (a) shows contours of interface temperature for different values of the normalized scattering and absorption coefficients.  This is the 2D version of Fig.~\ref{fig:3d-interface-temperature}.
	A normalized scattering of slightly above unity ($\overline{\sigma}_s$L$\sim$1.6) is taken, shown in Fig. \ref{fig:contour-interface} (b), to illustrate the effect of varying absorption on the interface temperature. Starting at a low value of absorption, typical of current plasma-sprayed GZO coating (purple), increasing absorption increases the interface temperature up to a maximum of about 30 K for $\overline{\kappa} L$$\sim$1, negatively impacting engine performance.  For further increases in absorption the interface temperature then decreases until a value (cyan), shown by the intercept of the dotted line, it again has a value corresponding to that of the current coating. Further increases in absorption, then lead to decreases in the coating/alloy interface as indicated by the descending arrow towards the opaque limit.  The decreasing interface temperatures would be beneficial for increasing engine efficiency by decreasing the necessary internal cooling.
	Also as shown in the low absorption regime, $\overline{\kappa}$ L$<$1, the interface temperature has a very strong dependence on the value assumed for the interfacial emissivity. Following previous studies, this value has been assumed to be $\epsilon_{int}$=0.30 but there are no direct measurements of this parameter.
    For high values of interfacial emissivity the radiation is absorbed near the interface region, thus, the temperature rises.
	
\begin{figure}[!htbp]
		\centering
		\includegraphics[width=1\columnwidth]{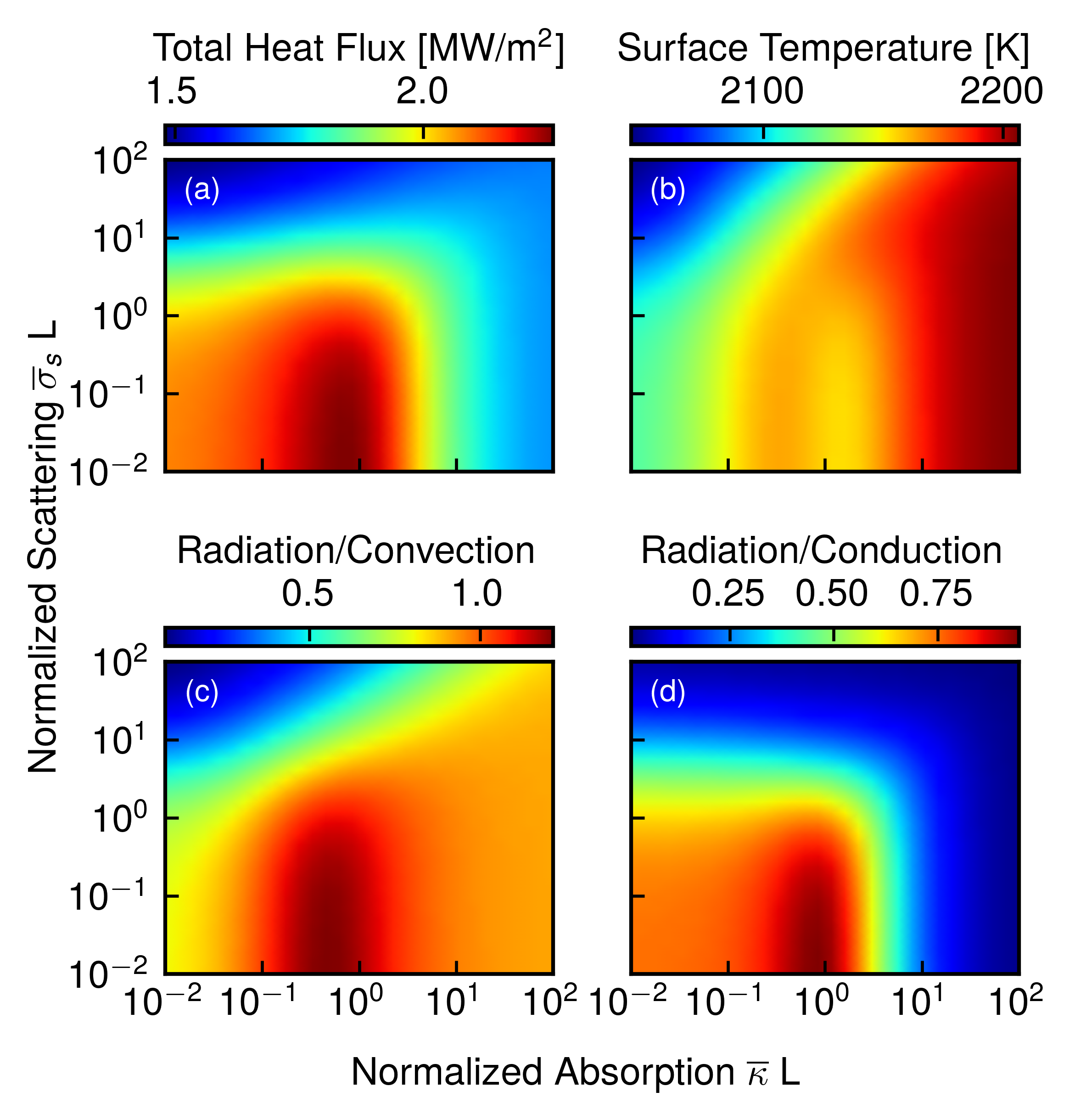}
		\caption{Contours of (a) total heat flux, (b) surface temperature, (c) the ratio of wavelength integrated net radiation energy to the convective heat transfer that passes through the surface and into the coating and (d) the ratio of net radiative energy to the conductive heat transfer at the interface.}
		\label{fig:contour-surface-interface-various}
	\end{figure}

	The total heat flux at the coating-alloy interface shown in Fig. \ref{fig:contour-surface-interface-various} (a) shows the total (conduction and radiation) heat flux contours as a function of normalized scattering and absorption lengths.  
    This has the same functional form on the scattering and absorption coefficients as the interface temperature shown above in Fig. \ref{fig:contour-interface} because of the steady state assumption that the total flux is constant through the thickness of the coating. In contrast, the surface temperature, see Fig. \ref{fig:contour-surface-interface-various} (b), does not exhibit the same variation since the thermal energy is preferentially absorbed in the surface region and is a maximum at the highest values of the normalized absorption length.
    
    Because of the differences in the surface and interface temperatures it is also of interest to compare the ratio of the net radiative heat flux to the convective flux, evaluated at the surface, and the ratio of the radiative to conductive components of the heat flux at the interface. These are shown in Fig. \ref{fig:contour-surface-interface-various} (c) and (d).

    The ratio of the radiative flux to the convective flux at the coating surface is given by the expression:
   
	\begin{equation}\label{totalradiationenergypassingthroughsurface}
		\frac{Q^{\prime \prime}_{\text{o, rad, total}}}{Q^{\prime \prime}_{\text{conv}}}
		= \frac{\splitdfrac{\int_{\lambda = 0}^{\lambda_\text{cut-off}} \dot{q}^{\prime \prime}_{\text{rad, }\lambda, x=0}  \,\mathrm{d} \lambda + (1 - \rho_o) \cdot}{ \int_{\lambda_{\text{cut-off}}}^{\infty} \left[ E_{\lambda b} (T_{\text{surr}}) - E_{\lambda b} (T_{\text{x=0}}) \right] \,\mathrm{d} \lambda}}{h_{\text{g}} \left(T_{\text{g}}-T_{\text{x=0}}\right)}
	\end{equation}
 
	As described above, current plasma-sprayed GZO coatings have normalized scattering and normalized absorption of about $\overline{\sigma}_s L$$\sim$1 and $\overline{\kappa} L $$\sim$10$^{-2}$, respectively. For these values, the total net radiative energy passing through the surface is commensurate to the convective heat flux. However this ratio decreases for more strongly absorbing materials, for a fixed coating thickness, primarily because more of the radiative energy is absorbed closer to the surface. This increases the surface temperature which, in turn, decreases the gas-coating surface temperature differential and, consequently, the heat convected into the coating. 
	
	The ratio of the radiative to conductive fluxes that reaches the interface can be computed from the expression: 
 
	\begin{equation}\label{radiation/conduction}
		\frac{Q^{\prime \prime}_{\text{int, rad, total}}}{Q^{\prime \prime}_{\text{int, cond}}} = \frac{\int_{\lambda = 0}^{\lambda_\text{cut-off}} \dot{q}^{\prime \prime}_{\text{rad, }\lambda,\text{ }x=L_1} \,\mathrm{d} \lambda}{\frac{k_1}{L_1} \left[T_{x=0} - T_{x=L_1}\right]}
	\end{equation}
 
	The radiative energy here includes only terms in the spectral translucent region up to the cut-off frequency. The radiation-to-conduction ratio at the interface exhibits similar characteristics to the interface temperature and the total heat flux shown in Fig. \ref{fig:contour-interface} (a) and \ref{fig:contour-surface-interface-various} (a), respectively.
 
	One of the striking features of the radiation/conduction ratio, see Eq. \ref{radiation/conduction}, shown in Fig.~\ref{fig:contour-surface-interface-various} (d) is the sharp drop at $\overline{\kappa} L$$>$1, which for  $\overline{\sigma}_s L$$\sim$1. This is attributed the greater absorption of radiation and consequently less thermal radiation reaching the coating/alloy interface. In turn, more of the thermal radiation is converted to phonons closer to the surface and contributing more to the thermal resistance.

\section{Spectrally-selective absorbers of thermal radiation}

    Having discussed the effects of absorption and scattering in terms of normalized wavelength independent parameters, this section presents the results for a gadolinium zirconate material that has been doped by different concentrations of Yb$^{3+}$, an ion that absorbs over a narrow spectral range due to thermal excitation of inner shell 4f electrons in the Yb ion. Then, in the following sub-section, the effects of a broader band absorption are considered.

	\subsection{Rare-earth element doped gadolinium zirconate }
	
		As shown in Fig.~\ref{fig:optical-gap-blackbody}, gadolinium zirconate,  \textit{i.e.} Gd$_{2}$Zr$_2$O$_7$, exhibits little or no adsorption over the thermal radiation window. However, this material exhibits complete solid solution of many rare-earth ions which can substitute in place of the Gd$^{3+}$ ion, without altering its crystal structure but absorbing at specific wavelengths dependents on the rare-earth ion replacing the Gd$^{3+}$. One such substitution for the Gd$^{3+}$ ion is the Yb$^{3+}$ ion, which produces optical absorption between 0.9 and 1.0 \si{\micro\metre}. Compositions all the way from Gd$_{2}$Zr$_2$O$_7$ to Yb$_{2}$Zr$_2$O$_7$ can be produced, and denoted by (Gd$_{1-x}$Yb$_{x}$)$_2$Zr$_2$O$_7$, where $x$ is the concentration of Yb$^{3+}$ ions replacing the Gd$^{3+}$ ions.

\begin{figure}[!htbp]
			\centering
			\includegraphics[width=1\columnwidth]{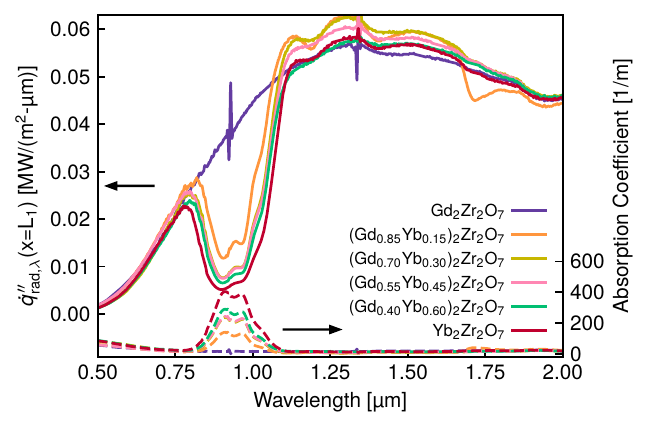}
			\caption{Net radiative spectral heat flux at the coating-substrate interface (left) and absorption coefficient (right) for different compositions of (Gd$_{1-x}$Yb$_{x}$)$_2$Zr$_2$O$_7$, where $0\leq x \leq 1.$}
			\label{fig:spectral-flux-absorption-Yb-compositions}
		\end{figure}

		The effect of the Yb$^{3+}$ absorption is shown in Fig.~\ref{fig:spectral-flux-absorption-Yb-compositions} which is a superposition of the interfacial net radiative heat fluxes (left, solid) and the absorption spectra (right, dashed) as a function of Yb concentration across a 500 \si{\micro\metre} solid solution (Gd$_{1-x}$Yb$_{x}$)$_2$Zr$_2$O$_7$ series. The experimental data exhibit a relatively broad absorption doublet centered at about 0.95 \si{\micro\metre} whose intensity increases in proportion to the Yb$^{3+}$ concentration up to the Yb$_2$Zr$_2$O$_7$ end member of the solid solution.  Using this spectral absorption data, the net radiative heat flux at the coating/alloy interface is computed to decrease over the same spectral region and does so in proportion to the Yb$^{3+}$ concentration.
		
\begin{figure}[!htbp]
			\centering
			\includegraphics[width=1\columnwidth]{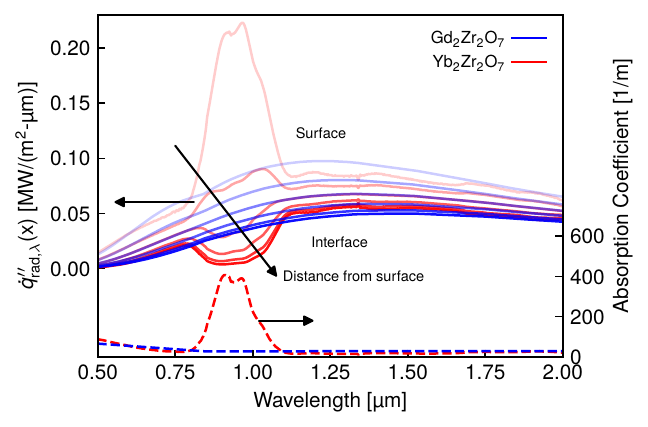}
			\caption{Comparison of the net radiative spectral heat flux distributions at different depths through the coating thickness for Gd$_{2}$Zr$_2$O$_7$ and Yb$_{2}$Zr$_2$O$_7$. The surface and the coating-alloy substrate interface are indicated by the most faded and most opaque curve, respectively.}
			\label{fig:spectral-flux-distribution}
		\end{figure}

		Figure \ref{fig:spectral-flux-distribution} shows the net radiative spectral heat flux distributions (left) at different depths through the 500 \si{\micro\metre} thickness for two coatings, Gd$_{2}$Zr$_2$O$_7$ (blue), which exhibits little or no absorption, and Yb$_{2}$Zr$_2$O$_7$ (red), that exhibits the absorption shown in the figure (right). 
		The coating surface and the coating-substrate interface are indicated by the most faded and most opaque curve, respectively.
		The higher absorption coefficient of Yb$_{2}$Zr$_2$O$_7$ has caused part of the radiative flux to be absorbed closer to the coating surface which, in turn, decreases the radiative energy reaching the coating-substrate interface relative to that when the undoped Gd$_{2}$Zr$_2$O$_7$ is used.
		This spectrally-selective heat transfer mechanism not only has impact on the thermal radiation, but it also lowers the convection heat transfer as was described above. Further discussion is provided in the text associated with Fig.~\ref{fig:contour-surface-interface-various}.
		
\begin{figure*}[!htbp]
			\centering
			\includegraphics[width=1\textwidth]{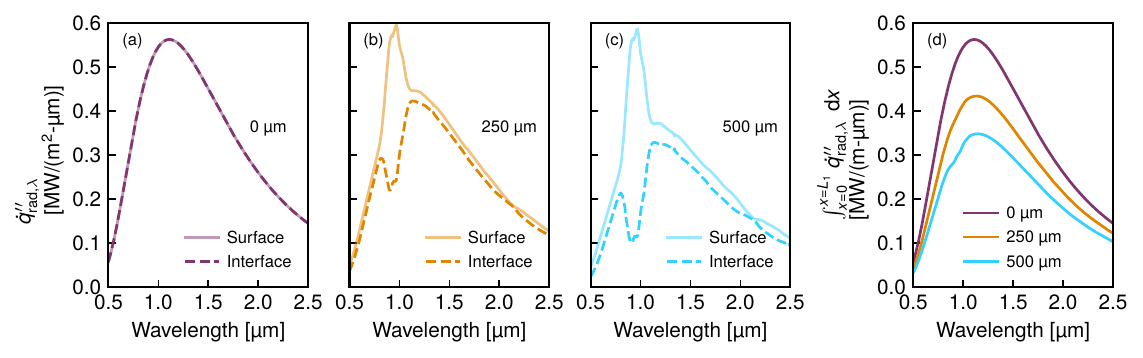}
			\caption{Net radiative spectral heat flux at locations (a) 0 \si{\micro\metre}, (b) 250 \si{\micro\metre} and (c) 500 \si{\micro\metre} shown at the coating surface (solid) and coating-substrate interface (dashed) for a Yb$_{2}$Zr$_2$O$_7$ material. (d) Net radiative spectral heat flux integrated up to the interface location x=L$_1$ for all the above coating thicknesses.}
			\label{fig:spectral-flux-interface-thickness-sweep}
		\end{figure*}

		A parametric study was carried out to explore the thickness effects on the thermal radiation for the Yb$_{2}$Zr$_2$O$_7$ coating material, as shown in Fig.~\ref{fig:spectral-flux-interface-thickness-sweep} (a, b, c, d), using the absorption data shown in Fig.~\ref{fig:spectral-flux-distribution}.
		These figures show the net radiative spectral heat flux at the coating surface (solid) and interface (dashed) for three different coating thicknesses: (a) 0 \si{\micro\metre}, (b) 250 \si{\micro\metre} and (c) 500 \si{\micro\metre}.
		Increasing the coating thickness the effect is similar to that shown above for absorption: more radiation is deposited in the surface relative to the interface.
		It is worth noting that the effect of spectrally selective absorption of Yb$_{2}$Zr$_2$O$_7$ is negligible even at 500 \si{\micro\metre}, as shown from the cumulative net radiation spectral heat flux integrated from x=0 to x=L$_1$ in Fig.~\ref{fig:spectral-flux-interface-thickness-sweep} (d).
        When the data for the Yb$_{2}$Zr$_2$O$_7$ material is superimposed on the curve in Fig.~\ref{fig:contour-interface} (b), it is seen that the Yb$^{3+}$ absorption has negligible effect on the temperature at the coating-substrate interface. This suggests that an oxide material with significantly greater absorption is required to lower the interface temperature for a gas-phase radiating as a black body. 
		
	\subsection{Broad Band Absorbers}
	
		In this section, we consider a more general, less specific form of broad band absorption in the same spirit as a gray body over a range of wavelengths is sometimes considered in radiative heat transfer problems. 
 
        \begin{figure}[!htbp]
			\centering
			\includegraphics[width=1\columnwidth]{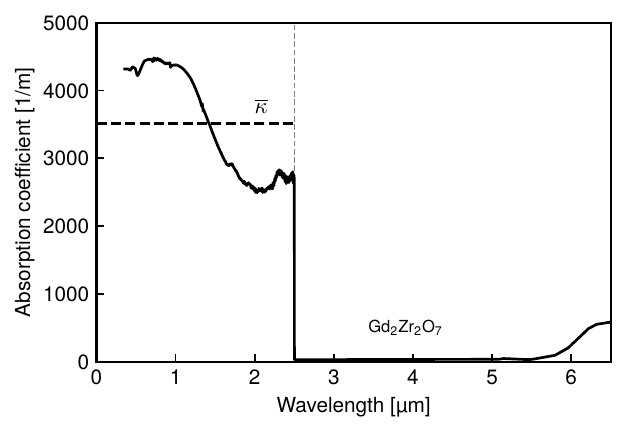}
			\caption{Absorption coefficient spectra of black GZO from 0.4 to 2.5 \si{\micro\meter} range. The absorption coefficients of Gd${_2}$Zr${_2}$O$_{7}$, see Fig.~\ref{fig:optical-gap-blackbody} were adapted for the remaining spectral range. The horizontal dashed line is the average, wavelength-integrated absorption coefficient.}
			\label{fig:absorption-coeff-black-alumina}
		\end{figure}
  
        One material that exhibits a broad absorption is a ``black'' alumina formed by doping polycrystalline alumina, which is otherwise translucent, with both TiO$_2$ and Fe$_2$O$_3$ (to be published). It's absorption spectrum is shown in Fig.~\ref{fig:absorption-coeff-black-alumina} up to 2.5 \si{\micro \meter} deduced from reflectance measurements of an about 100 \si{\micro\meter} sample using the Kubelka-Munk method. As has already been remarked, nearly 75 \% of the black body radiative flux at 2500 K is emitted in this spectral region. To compute the effects of such an absorption spectrum on the radiative heat transport in GZO, we use all the other properties of the GZO and simply use the ``black alumina'' absorption spectrum in place of the absorption characteristics of Gd${_2}$Zr${_2}$O$_{7}$ used in the rest of this contribution. This hypothetical material is referred to here as ``black GZO''. The absorption is not only broader than that due to the electronic transitions in the Yb$^{3+}$ ions, shown in Fig. \ref{fig:spectral-flux-distribution}, and considered in the previous sub-section but also the integrated absorption over the measured spectral band is nearly sixty-times larger.
        For the rest of the translucent band, up to $\lambda_{\text{cut-off}}$, the value of the absorption of Gd${_2}$Zr${_2}$O$_{7}$ is used, shown in Fig.~\ref{fig:optical-gap-blackbody}.
  
\begin{figure*}[!htbp]
			\centering
			\includegraphics[width=1\textwidth]{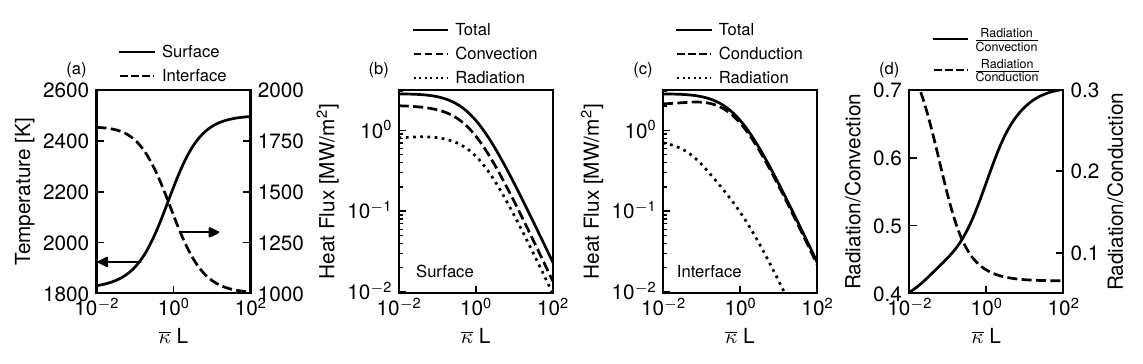}
			\caption{Thermal performance characteristics of ``black GZO'' as a function of the normalized absorption length $\overline{\kappa}$ L. (a) Surface (solid) and interface (dashed) temperatures are shown on the left and right vertical axes, respectively. (b) The total heat flux (solid), the convective heat flux contribution (dashed) and wavelength-integrated net radiative flux contribution (dotted) evaluated at the surface. (c) The total heat flux (solid), conductive heat flux (dashed) and wavelength-integrated net radiative flux (dotted) evaluated at the coating/alloy interface by assuming an energy balance at the interface. (d) Ratio of total net radiative energy passing through the surface to the convective heat flux (solid) and the ratio of net radiative energy reaching the interface to the energy conducted to the interface (dashed). Coating thickness: 500 \si{\micro \meter} }
			\label{fig:qT-thickness-sweep-two-materials}
		\end{figure*}

        Fig.~\ref{fig:qT-thickness-sweep-two-materials} shows results for the black GZO material using the absorption coefficient of Fig.~\ref{fig:absorption-coeff-black-alumina}.
        The spectral analysis utilized the wavelength-dependent absorption coefficient (solid) for the full translucent band.
        The results were normalized with respect to $\overline{\kappa}$L by fixing the wavelength-integrated absorption coefficient (dashed) and varying the coating thickness, $L$.
	
		A thermal performance analysis of the ``black GZO'' is shown in Fig.~\ref{fig:qT-thickness-sweep-two-materials}.
		For increasing normalized absorption length the surface (left) and interface (right) temperatures are increased and decreased, respectively, see Fig.~\ref{fig:qT-thickness-sweep-two-materials} (a).
		This absorption increase causes more of the radiative heating to be absorbed near the surface and correspondingly less to reach the interface. In turn, the increase in surface temperature decreases the difference with the gas temperature and decreases convection. This effect can be seen from the energy balance at x=0 in Fig.~\ref{fig:qT-thickness-sweep-two-materials} (b).
		Furthermore, the wavelength-integrated net radiative energy also decreases for increasing thickness.
		Consequently, the total heat flux follows similar behavior as its components.
		Because the total thermal energy is conserved and steady-state is assumed, the same heat flux will reach the interface. This is revealed by an energy balance performed at x=L$_1$, shown in Fig.~\ref{fig:qT-thickness-sweep-two-materials} (c).
		At very small values of normalized absorption the relative contribution of radiation to the total heat flux is comparable between the surface and interface, because the material is, apart from scattering, transparent.
		However, the total net radiation energy becomes significantly smaller for large values of normalized absorption as compared to the total heat flux or radiation at the surface.
		Lastly, Fig.~\ref{fig:qT-thickness-sweep-two-materials} (d) shows that the surface radiation becomes comparable to convection, and the interfacial radiation becomes insignificant as compared to heat conduction for increasing values of normalized absorption.
  
\section{Discussion}

    To investigate the possible effects of radiative heating on the heat fluxes and temperatures, we have used GZO as a reference material for three main reasons: it is one of the commercial coating materials used in current engines and can sustain much higher temperatures; there is reasonable property data base available in the open literature; and we have demonstrated that it is possible to alter it's optical absorption by substitutional doping with rare-earth ions without affecting its thermal conductivity~\cite{Champagne2023}. We have also used values for some of the key heat transfer parameters, such as the appropriate convective heat transfer coefficients, that have been used in previous studies of coatings~\cite{siegel1998analysis,flamant2019opportunities}. We also point out that the thermal (phonon) conductivities, $k_1$ in Eq. \eqref{energybalancefunctionofx}, of refractory oxides above about 1500 K are independent of temperature so the temperature dependencies of the heat fluxes and temperatures calculated primarily depends on the propagation of thermal radiation through the coatings. 
    
    One of the key parameters in the thermal-mechanical design of thermal barrier coatings, as well as for materials selection criteria, is the coating thickness. For instance, for rotary components, the coating thickness is one that maximizes the temperature drop through the coating, while minimizing the additional weight associated with the coating and the centrifugal forces acting on the coating as well as the alloy component. For a stationary component, such as a combustor liner, which is not subject to centrifugal forces, the coating thickness can be greater, sometimes well in excess of 1000 \si{\micro \meter}. When thermal conduction alone determines heat transfer through the coating, then the thermal resistance increases linearly with the coating thickness. For conditions under which thermal radiation may be significant, the length scale is different and is determined by the extinction coefficient. For this reason, we have used the coating thickness, L, to normalize the absorption and scattering coefficients to identify conditions under which thermal radiation may be significant in translucent coatings. The effects of thickness on the heat transfer through a coating are clearly shown in the Fig.~\ref{fig:qT-thickness-sweep-two-materials}. The most appreciable changes in heat fluxes at both the surface and coating/alloy interface occur when $\overline{\kappa}$ L$>$ $e^{-1}$.  To put this value in context, it is useful to consider current coating thicknesses, which range from about 150$\cdot$10${^{-6}}$ \si{\meter}, for a first row blade, and 10$^{-2}$ \si{\meter} for a combustor liner.  Together these constraints not only indicate that absorption coefficients have to be larger when used for the thinner coating but also that a minimum value of the absorption coefficient larger than 10$^3$ to 10$^4$ m${^{-1}}$ is required for a thermal radiation barrier coating. More precise values depend on the thermal conductivity of the coating material, as well as on the boundary conditions. Similarly, the convective heat flux was found to have a strong dependence on the location where the radiative energy is absorbed. The greater the absorption and the closer to the surface is the radiative flux absorbed and correspondingly the higher the surface temperature.  In turn this decreases the gas-coating surface temperature differential and consequently the convective heat flow into the coating.
    
    Recognizing the growing importance of radiative heating in coatings a number of papers have recently been published on altering the optical absorption or emissivity of oxides  ~\cite{he2009high,dong2020thermal,zhao2019la0,qiu2022medium,chen2023near,huang2022opaque}. 
    One approach, exemplified by our studies presented in this work, is to incorporate into the crystal structure of a candidate coating material, ions with an electronic structure suited to absorb in the visible and near-infra-red. Another example is co-doping elements (Ca, Fe) or (Sr, Mn) in a La$_2$Ce$_2$O$_7$ material. However, increasing the concentration of the dopants in this material resulted to an increase in thermal conductivity as well as a reduction in both fracture toughness and thermal cycling lifetime~\cite{dong2020thermal}. Mixed rare-earth elements doping in oxides~\cite{wan2011order, zhao2019la0} also leads to enhanced infrared absorption and calcium-magnesium-alumina-silicate (CMAS) resistance~\cite{qiu2022medium}. In a variant of the doping strategy, Chen \emph{et al.}~\cite{chen2023near} recently showed that the absorption could be enhanced by incorporating into a translucent coating matrix phase a second phase and achieved a broadband infrared spectral emissivity (above 0.9) in a Ca-Cr co-doped Y$_3$NbO$_7$ without adversely increasing its thermal conductivity. Similarly, Huang \emph{et al.}~\cite{huang2022opaque} created a two phase coating of GdMnO$_3$ in Gd$_2$Zr$_2$O$_7$. Interestingly, the thermomechanical properties of these materials remained unchanged (up to 5\% GdMnO$_3$) and the wavelength-integrated absorption coefficient reached about 10$^{4}$ m$^{-1}$ up to 1 \si{\micro\meter} wavelength, before rapidly decreasing. This is closest to the absorption achievable with the ``black GZO'' but narrower in spectral width. Although none of these materials, except the La$_2$Ce$_2$O$_7$ material, has, to our knowledge, been deposited as coatings, the reports demonstrate that there is potential for varying the optical absorption properties of translucent oxides. However, as illustrated by the modeling performed in this work, large values of the product of the normalized absorption length and spectral absorption width will be required to significantly decrease radiative heat transport through coatings.

    From a materials selection perspective, however, it is generally desirable to keep both the coating surface temperature as well as the coating-substrate interface temperature as low as possible. High surface temperatures promote sintering and densification as both are thermally activated processes. Unfortunately, these densification processes are detrimental to coating life as they can lead to cracking and coarsening of porosity that reduces scattering ~\cite{limarga2009characterization} as well as reducing the volume fraction of porosity. The latter is important as porosity is needed for maintaining a low biaxial elastic modulus and low thermal conductivity. A low interface temperature is also desirable to minimize the oxidation rate of the underlying alloy. 
    

    A significant finding of these calculations is the sensitivity of the coating/alloy temperature to the value of the emissivity of the coating/alloy interface. This sensitivity is shown in Fig.~\ref{fig:contour-interface} and is particularly strong in the low absorption regime. As stated earlier, we have adopted the value $\epsilon_{int}$=0.30 used in previous studies but to our knowledge no actual measurements of this important parameter exist. Furthermore, as the interface evolves during service at high temperatures, for instance by oxidation and rumpling, it is a key parameter in calculating changes in radiative heat transfer with aging.
        
    In presenting this work it is recognized that one shortcoming of the continuum model is that the description of the scattering within the coating is too simple and doesn't properly represent scattering from microstructural features known to exist in coatings, such as pores and cracks, or for plasma-sprayed coatings, splat boundaries. These range in microstructural size, orientation and size distribution. These all affect the magnitude of optical scattering at different wavelengths and so need to be treated using Monte Carlo methods. Furthermore, because of their physical size, only a small number of each, except micron and sub-micron pores, can be in any thickness section ~\cite{flamant2019opportunities}. Nevertheless, the continuum model for scattering does indicate that there are conditions, $\overline{\sigma}_s L$$>$10, under which radiative scattering may be significant as shown in Fig.~\ref{fig:contour-interface}.

    Finally, we have also assumed, as have previous studies, that the hot gas radiates as a black body source. This may be a reasonable approximation to a combusted gas containing soot, but is probably not for combustion of jet fuels except under non-equilibrium conditions. Under clean burning conditions, the emissivity of the hot gas in the ``thermal radiation window'' will be smaller than that of a black body although not zero. The emissivity of the hot gas due to the presence of water vapor and CO$_2$ depends on both path length and pressure and can be calculated from the NIST Opensource Tool, OpenSC~\cite{Tam2019}. For instance, the emissivity due to pressure broadening at 50 bar is estimated to be $\sim$0.5-0.7 in the ``thermal radiation window''. Consequently, although the radiative fluxes, and the increases in the resulting local temperatures, are both expected to be lower than those computed in this work, they are expected to nonetheless be significant. Also, there are other related coating applications, such as at leading edges and hypersonic surfaces, where the temperature is so high that the radiation is from an excited plasma and the black body thermal radiation approximation is likely to be appropriate.

\section{Concluding Remarks}
	
    This paper has considered how optical absorption in translucent coatings, in the visible/infra-red region of the thermal radiation spectrum from a hot-gas, could affect the radiative heating of thermal barrier coatings and the additional internal cooling required to extract heat from alloy components.  The calculations, based on a spectral two-flux heat transfer model, indicate that there are combinations of radiative absorption, radiative scattering and coating thickness that can reduce both the coating/alloy interface temperature and the heat flux through the coating, and other combinations where they have little effect. These have been identified using parametric analysis using wavelength-independent values of the radiative absorption and scattering, normalized by the coating thickness. Two classes of spectrally selective absorbers were then considered, a narrow-band rare-earth zirconate, \textit{i.e.} Yb$_2$Zr$_2$O$_7$, and a broad band, \textit{i.e.} ``black GZO''.  The former is an example of substitutional doping that absorbs over a narrow spectral range, between about 0.9 and 1.0 \si{\micro\metre}, but because of the maximum attainable absorption coefficient does not exceed $\sim$500 m$^{-1}$ there is not an appreciably reduction in the radiative heating through a 500 \si{\micro\metre} thick coating. The latter type of material, ``black GZO'', is more effective since it exhibits a broader spectral absorption and an almost order of magnitude larger absorption coefficient. 
    
    Although the focus of this work has been on thermal barrier coatings for turbine blades, the results of this work apply directly to coatings on combustors and environmental barrier coatings (EBC) for ceramic matrix composites (CMC) as well as for materials for scramjets and hypersonic applications, where the temperatures may be even higher than considered here.

	Lastly, it is pertinent to mention that under very rapid changes in gas temperature, for instance, under flame-out (cold shock) or sudden increases in combustion temperature (up-shock), the thermo-mechanical stresses on the coating are expected to be very different than has been modeled for the thermal conduction controlled transient conditions considered previously ~\cite{sundaram2013influence, staroselsky2019influence, koutsakis2022fracture}. The thermal conduction propagation is by diffusion and so takes a characteristic time to propagate through the coating. This time is given by $\sim$$L^2/\alpha$ where L is the coating thickness and $\alpha$ is the thermal diffusivity. By contrast, the radiative shock propagates essentially instantaneously through the coating and can, depending on the properties of the material, cause an inversion of the temperature gradient in the coating. The consequences of this on the stress state in a coating as well as thermo-mechanical conditions for coating failure will be explored in a later contribution.

    

\section{Acknowledgments}
	This work was supported by the Office of Naval Research under grant N00014-21-1-2478. The authors are grateful to Victor K. Champagne for supplying the absorption coefficient data used in Fig.~\ref{fig:spectral-flux-absorption-Yb-compositions} and to Dr. Curt Johnson, General Electric (GE), for helpful discussions.

\section{Appendix}

	\begin{table}[!htbp]
		\centering
		\begin{tabular}{@{}ll@{}}
			\textbf{Nomenclature} & \\
			$\boldsymbol{\dot{q}^{\prime \prime}}$ & Heat Flux [Wm$^{-2}$] \\
            $\boldsymbol{G}$ & Incident radiation [Wm$^{-2}$m$^{-1}$] \\
            $\boldsymbol{T}$ & Temperature [K] \\
            $\boldsymbol{h}$ & Heat transfer coefficient [Wm$^{-2}$K$^{-1}$] \\
			$\boldsymbol{k}$ & Thermal conductivity [Wm$^{-1}$K$^{-1}$] \\
			$\boldsymbol{\lambda}$ & Wavelength [m] \\
			$\boldsymbol{n}$ & Refractive index [--] \\
            $\boldsymbol{L}$ & Thickness [m] \\
            $\boldsymbol{\kappa}$ & Absorption coefficient [m$^{-1}$] \\
            $\boldsymbol{\sigma_s}$ & Scattering coefficient [m$^{-1}$] \\
            $\boldsymbol{\beta}$ & Extinction coefficient [m$^{-1}$] \\
	        $\boldsymbol{\omega}$ & Scattering albedo [--] \\
            $\boldsymbol{E_{b}}$ & Blackbody emissive power [Wm$^{-2}$] \\
            $\boldsymbol{\epsilon}$ & Emissivity [--] \\
            $\boldsymbol{\rho}$ & Reflectivity [--] \\
            $\boldsymbol{\sigma}$ & Stefan–Boltzmann constant [Wm$^{-2}$K$^{-1}$] \\
         
            & \\
			
			Subscripts & \\

            g & gas convection from combustion side \\
            c & gas convection from cooling side \\
			cond & conduction \\
			conv & convection \\
			rad & radiation \\
			w & wall \\
            m & metal \\
			surr & surroundings \\
            i & inner or incident\\
            int & interface \\
            o & outer \\
			$x$ & $x$-direction \\

            & \\
            
			Superscripts & \\
						
			+ & positive direction (right) \\
			- & negative direction (left) \\
			
			& \\
			
			Abbreviations & \\
			
			\textbf{TBC} & Thermal Barrier Coating \\
			\textbf{EBC} & Environmental Barrier Coating \\
			\textbf{YSZ} & Yttria Stabilized Zirconia \\
            \textbf{GZO} &  Gadolinium Zirconate \\
            \textbf{CMC} & Ceramic Matrix Composites \\
            \textbf{CMAS} & Calcium-magnesium-alumina-silicate \\ 
            \textbf{UV-VIS} & Ultra Violet and Visible spectrum \\
            \textbf{EB-PVD} & Electro-Beam Physical Vapor Deposition \\            
		\end{tabular}
	\end{table}

\bibliography{./library.bib}

\appendix

\end{document}